\documentstyle[psfig]{l-aa}

\begin{document}

   \thesaurus{06         
              (03.11.1;  
               16.06.1;  
               19.06.1;  
               19.37.1;  
               19.53.1;  
               19.63.1)} 
   \title{Near-infrared photometry of the young open clusters NGC~1893 and
Berkeley~86 \thanks{Based on observations taken at TIRGO}}

   \author{Antonella Vallenari 
\inst{1}, Andrea Richichi\inst{2}, Giovanni Carraro\inst{3}, L\'eo Girardi\inst{3}
          }

   \offprints{Antonella Vallenari ({\tt vallenari@pd.astro.it})}

   \institute{Padova Astronomical Observatory, vicolo Osservatorio 
          5, I-35122, Padova, Italy
        \and
        Arcetri Astrophysical Observatory, Largo E. Fermi
        I-50110, Firenze, Italy
         \and
        Department of Astronomy, Padova University,
	vicolo dell'Osservatorio 5, I-35122, Padova, Italy\\
        e-mail: {\tt 
vallenari\char64pd.astro.it,richichi\char64arcetri.astro.it,carraro,lgirardi\char64pd.astro.it,
}
             }

   \date{Received May 1999 ; accepted August 1999 }

   \maketitle

   \markboth{Vallenari et al}{Infrared photometry}

   \begin{abstract}
We present photometry in the J and K near-infrared bands for
two regions centered on the young open clusters  
NGC~1893 and Berkeley~86. We study 700 stars down to K = 17 in the field
of
NGC~1893, 
and about 2000 stars in the field of Berkeley~86 down to K$\sim$ 16.5,
 for which near-infrared
photometry was insofar not available. 
Coupling J-K data with UBV photometry taken from literature,
we produce reddening corrected colour-magnitude diagrams.
 We find that our data are consistent with previous
determinations: the clusters are roughly coeval
with an age between 4 and 6 million years. 
The mean reddening (measured as E(J-K)) values
turn out to be 0.35 and 0.50 for NGC~1893 and Berkeley~86,
respectively.

Using colour-colour plots we discuss the presence of
candidate pre-main sequence stars showing infrared excess.
Candidates are found in both cluster regions, confirming the
young age of these clusters.

      \keywords{Photometry: infrared -- Open Clusters --\\
                NGC~1893: individual --
                Berkeley~86: individual
               }
   \end{abstract}

%

\section{Introduction}

Open clusters and associations in the Galactic disk are the best tools
 to investigate the distribution of population inside the
disk and the spiral arms. 
We have undertaken a
 project to obtain deep near-infrared photometry
of Galactic open clusters, aimed at deriving age, age spread, reddening,
distance.
About 10 clusters have already been observed.
 In the first paper, the ages of the old open clusters Berkeley~18 and
 Berkeley~17
have been derived and compared with the age of the Galactic disk (Carraro
et al 1999).
In this paper we present J and K photometry for two young open
clusters: NGC~1893 and Berkeley~86, whose fundamental properties
are listed in Table~1. Two forthcoming papers will  discuss   the age of
King~5 and IC~166, NGC~7789 respectively. \\

NGC~1893 is a very young cluster involved in the bright diffuse
nebulosity IC~410, associated with two pennant nebulae, Shain and Gaze
129 and 130,
and obscured by several conspicuous dust clouds.\\
UBV photometry of NGC~1893 has been carried out by Cuffey (1973)
and Massey et al (1995).
 More than 39 members are
bright early spectral type stars, responsible for the photoionization
of the IC~410 nebula.\\
Tapia et al (1991, TCER hereafter)  perform near-infrared and Str\"omgren
photometry for 50 stars down to K = 12.00. They
estimate the age of the cluster to be $4\times 10^{6}$ yr, derive
the distance modulus  $(M-m)_{0}=13.18 \pm 0.11$, and the reddening A$_v$=1.68.
Str\"omgren photometry for 50 stars in the field of NGC~1893
has also been reported by Fitzsimmons (1993), who confirms the
distance and age found by TCER.\\

\begin{figure*}
\centerline{\psfig{file=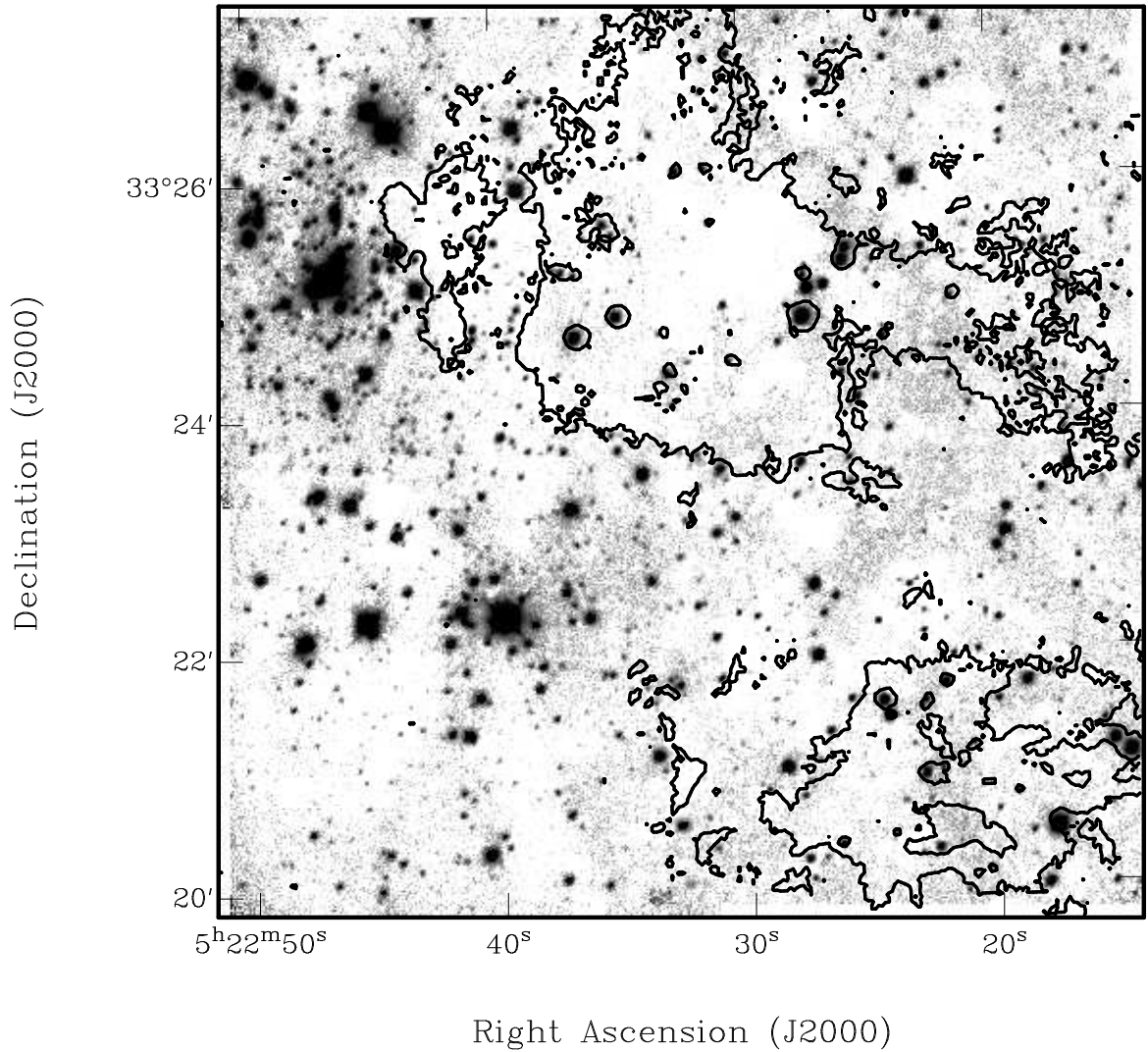,width=16cm,height=16cm}}
\caption{\label{image1}  Image in J band covering the studied region of
NGC~1893.  North is on the top, east at the left.
 The contours  represent
 the dark clumps visible in the DSS image as explained in the text.}
\end{figure*}

\begin{figure*}
\centerline{\psfig{file=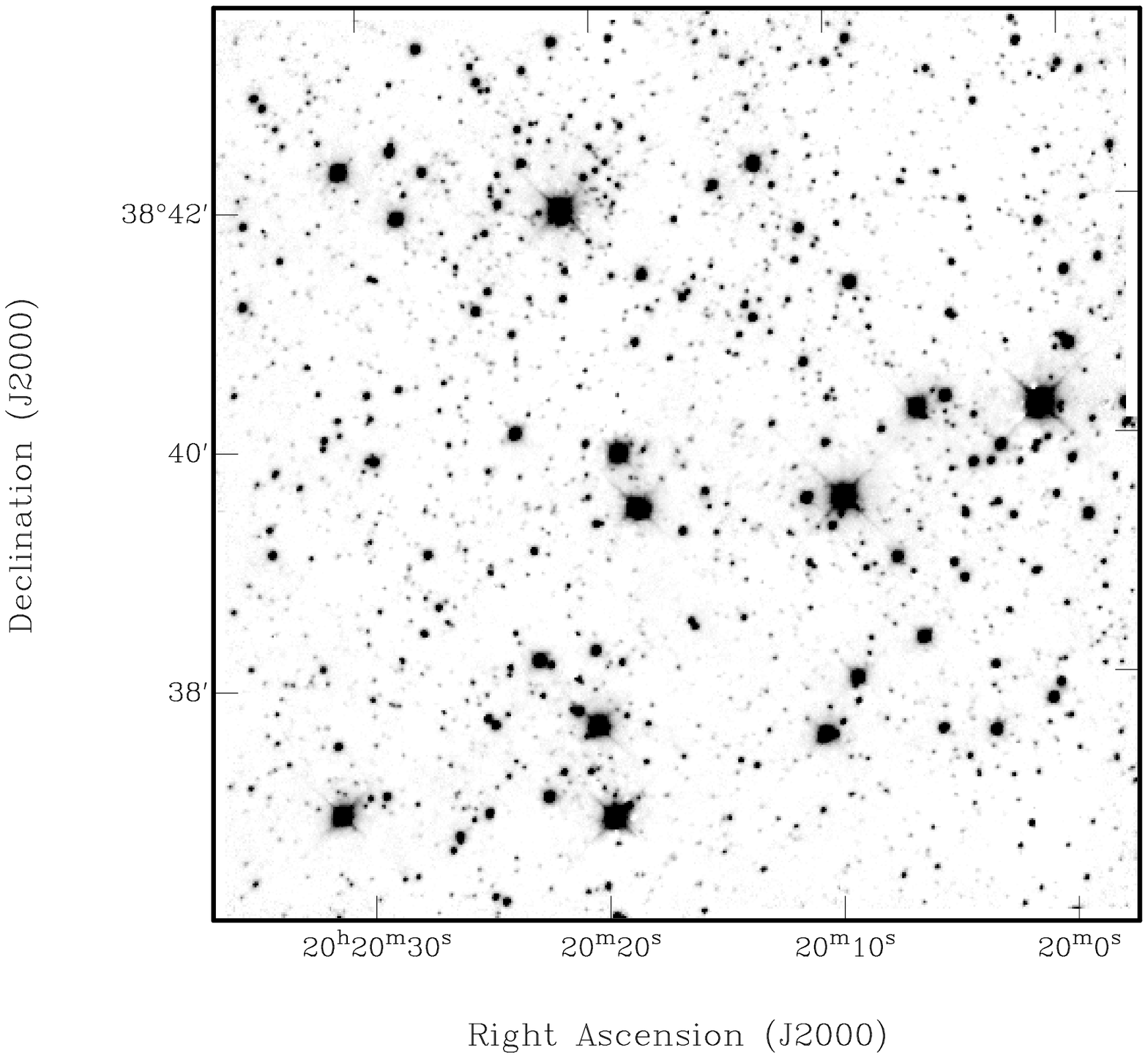,width=16cm,height=16cm}}
\caption{\label{image2} Image in J band covering the studied region of
Berkeley~86. North is on the top, east at the left.}
\end{figure*}

Berkeley~86  is another young open cluster, located inside the OB
association Cyg OB 1, and obscured by a foreground dust cloud.
Optical photometry was carried out by Forbes et al (1992) in the UBV
pass-bands, and in the UBVRI bands by Deeg \& Ninkov (1996) and Massey et al
(1995).
They find that the cluster has an age of $5 \times 10^{6}$ yr,  is
moderately reddened (E(B-V) $\approx 1.0$) and has an initial mass
function close to the Salpeter one.  Berkeley~86 hosts the famous eclipsing
binary system V444 Cygni.\\
Str\"omgren photometry has also been obtained by Delgado et al (1997)
down to V = 19, who derive a slightly older age ($8.5 \times 10^{6}$ yr)
and a distance modulus of $(M-m)_{0} = 11.1$. \\
No infrared photometry exists for this cluster.\\

In this paper we present near-infrared photometry for these two
clusters, down to the limiting magnitude  of K $\sim 16.5-17.0$.
 In Section~2 we detail the observations and data reduction process.
 In Section~3 and 4 we discuss in turn  for NGC~1893 and Berkeley~86 
clusters the derived colour-magnitude diagrams and the detection of pre-main sequence  candidates.
We summarizes
 our results in Section~5.

\begin{table*}
\tabcolsep 0.10truecm
\caption{Basic parameters of the studied clusters.}
\begin{tabular}{lccccccc} 
\hline
\hline
\multicolumn{1}{c}{Cluster} &
\multicolumn{1}{c}{$l$} &
\multicolumn{1}{c}{$b$} &
\multicolumn{1}{c}{Diameter} &
\multicolumn{1}{c}{Age} &
\multicolumn{1}{c}{A$_V$} &
\multicolumn{1}{c}{$(M-m)_{0}$}&
\multicolumn{1}{c}{Reference} 
\\
 &$^{o}$ &$^{o}$ &($\prime$)&Myr&&& \\
\hline
NGC~1893       & 173.59 & -1.70 & 13 & 4&1.68&13.18& Tapia et al (1991)\\
Berkeley~86    &  76.66 &  1.26 &  6 &5&3.1&11.1& Massey et al (1995)\\
&&&&8.5&&& Delgado et al (1997) \\
\hline\hline
\end{tabular}
\end{table*}

\section{Observations and data reduction}
	
J (1.2 $\mu$m) and K (2.2 $\mu$m) photometry of the two clusters was obtained 
with 1.5m Gornergrat Infrared Telescope (TIRGO) 
equipped with Arcetri Near Infrared Camera (ARNICA)
in October 1997.
ARNICA relies on a NICMOS3 256$\times$ 256
pixel array (gain=20 e$^-/$ADU, read-out noise=50 e$^-$
 angular scale =1$\arcsec/$pixel, and 4$^\prime \times$ 4$^\prime $  field of view). 
Through each filter 4 partially overlapping images of
each cluster were obtained,
 covering a total field of view of about 8 $^\prime \times$ 8 $^\prime$ ,
in short exposures to avoid sky saturation. 
The observed field of NGC~1893 is covering approximately the SW  quadrant
of the region studied by Tapia et al (1991). Only about $1/4$ of the
total cluster area has been covered by our observations. The field
was chosen to study the stellar content
 of the dark clumps in NGC~1893.
The observed field of   Berkeley~86 includes the vast majority of this cluster.
 Berkeley~86 was
studied by Deeg \& Ninkov (1996) who derive U,B,V,R,I photometry
for stars in a slightly smaller field of  6$\arcmin \times$6$\arcmin$.
The log-book of the observations is given in Table~2 were the total exposure times are given. 
  The nights were photometric
with a seeing of 1$\arcsec$-1.5$\arcsec$. 
Figs. \ref{image1} and \ref{image2} present the mosaics of the 4 frames obtained per cluster
in J passband.

The data are reduced  subtracting from each image a linear combination of the
corresponding  skies and dividing the results by flat fields
taken on twilight sky. We make use of 
the Arnica package (Hunt et al 1994)
in IRAF.  Daophot II is used to perform photometry.

The conversion of the
instrumental magnitude j and k to the standard
J, K is made using stellar fields of standard stars taken
by   Hunt et   al (1998) list. We point out that the
JHK Tirgo system is found by Hunt et al (1998) to be
 consistent with the UKIRT system
as described by Casali \& Hawarden (1992).
 About 10 standard stars per night have been used.
The  relations  per 1 sec exposure time are:
 
\begin{equation}
J  = j+19.51 \\
\end{equation}

\begin{equation}
K  = k+18.94
\end{equation}

\begin{table*}
\caption[ ]{ Observation Log-Book. The coordinates listed below refer to
the center of the mosaic.}
\begin{tabular}{c|c|c|c|c|c}
\hline
\hline
Cluster               &$\alpha$    &$\delta$  & Date& 
\multicolumn{2}{c}
{Exposure Times (sec)} \\ 
                           &(2000)      &(2000) &    &J&K\\
\hline
NGC 1893    &55 22 34  & 33 23 29 & Oct, 22, 1997& 720 & 660 \\
Berkeley 86 &20 20 18 & 38 39 27 & Oct, 25, 1997& 420 & 420 \\
            &          &          & Oct, 26, 1997& 480 & 420 \\     
\hline
\hline
\end{tabular}
\end{table*}

\noindent
with standard deviation of the zero points of 0.03  mag for the J 
and 0.04 for the K magnitude. This error is only due to the linear
interpolation of the standard stars. The calibration uncertainty
is dominated by the error due to the correction from aperture photometry
to PSF fitting magnitude. Taking it into account,
 we estimate that the total error
on the calibration is about 0.1 mag both in J and K pass-bands.
The standard stars used for the calibration do  not cover the entire
colour range of the data, because of the lack of stars redder than
(J-K)$ \sim 0.8$. From our data, no colour term is found for K mag,
whereas we cannot exclude 
it  for the J magnitude.
The limiting magnitudes are K $\sim$ 17 and 16.5 for NGC~1893
and Berkeley~86 respectively.

The data tables will be published electronically
at the Centre des Donnes
Stellaires  of Strasbourg.

\begin{figure}
\centerline{\psfig{file=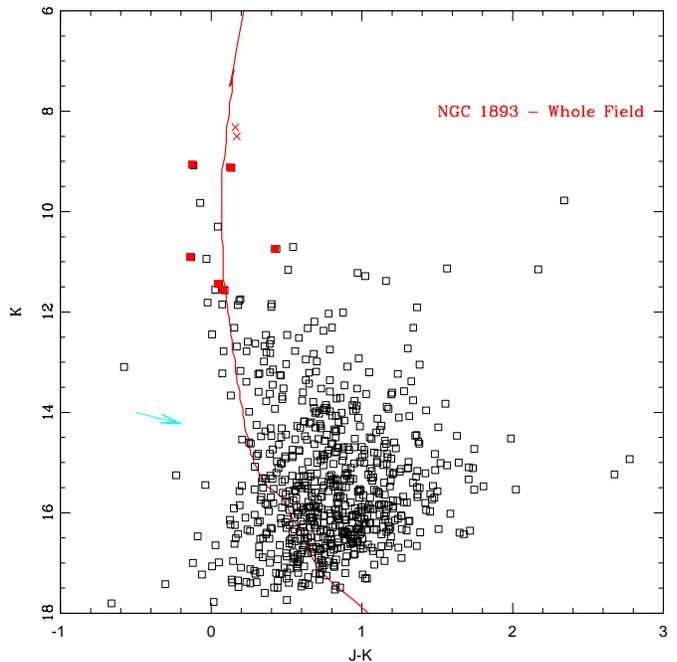 ,width=9cm,height=9cm}}
\caption{\label{n1893iso} The  CMD of the whole observed region
inside NGC~1893  
with isochrone from Padova models having  an age of 4 Myr and E(J-K)=0.35.
The arrow indicates the reddening vector corresponding to  E(J-K)=0.35.
Solid squares indicate bona fide members on the basis of Tapia et al
spectroscopy. Crosses shows the bona fide members brighter
than K=9 not included in our observed field taken from Tapia et al
photometry.
}
\end{figure}

\begin{figure}
\centerline{\psfig{file=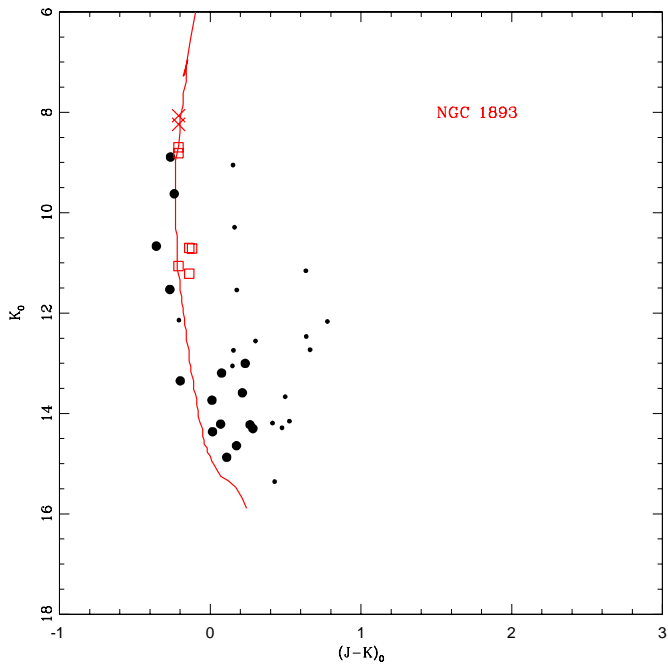,width=9cm,height=9cm}}
\caption{\label{n1893isored}
De-reddened CMD for 38 stars in the field of NGC~1893, obtained as described in the text. An isochrone from Padova models having  an age of 4 Myr is plotted.
Solid circles represent candidate members, selected on the basis of the reddening. 
 Open squares indicate bona fide members on the basis of Tapia et al
spectroscopy. Crosses show  the bona fide members brighter
than K=9 not included in our observed field taken from Tapia et al
photometry.}
\end{figure}

\section{NGC~1893}
\subsection{The Color-Magnitude Diagrams}

The K-(J-K) CMD for NGC~1893 is shown in Fig.\ref{n1893iso}.

The cluster MS extends vertically between K=$ 8-8.5$ and 14, while at
fainter magnitudes it is contaminated by field stars.\\

We compare our photometry with
TCER one.
 Only 10 bright stars (down to K $\sim$ 13)
 in common  are found.

The comparison between our  photometry and TCER photometry 
gives:

\begin{eqnarray}
K \ - \ K_{TCER} \ = \ 0.013, \ \ \sigma=0.07 \\
(J-K)\ - \ (J-K)_{TCER} \ = \ 0.029, \ \ \sigma=0.08
\end{eqnarray}

\noindent
To investigate whether this JK CMD is consistent 
with Tapia et al cluster parameter determination,
we assume their determination of reddening and distance and
 we over-impose on the CMD an isochrone (taken from Padova models,
Bertelli et al 1994, Girardi et al 1999) of age of 4
million years (cf Fig.\ref{n1893iso}). Bona fide members  brighter than K=9
located outside the observed field, but included in Tapia et al
photometry are also taken into account. 
A good agreement with the CMD  is found.
A de-reddened CMD will be presented and discussed in the following Section.

\begin{figure}
\centerline{\psfig{file=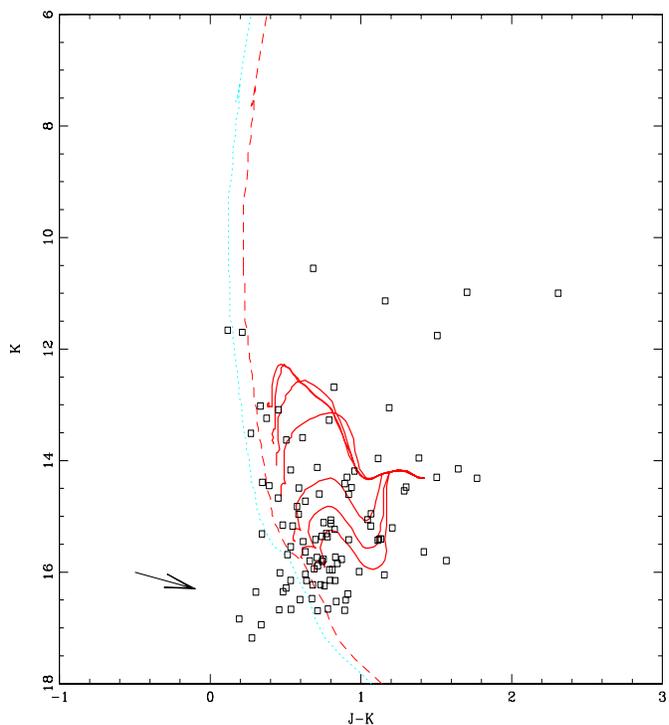,width=9cm,height=10cm}}
\caption{\label{cmd1893clump} CMD of the dark clump inside
NGC~1893, as discussed in the text, selected on error on the
magnitude $\sigma < 0.1$. Isochrones  corresponding to
an age of 4$\times 10^6$ yr,
 E(J-K)=0.55 and   E(J-K)=0.45 (long dashed line and short dashed line respectively)
are also shown, together with pre-main sequence
tracks by Bernasconi with E(J-K)=0.55  for stellar
masses 7,5,4,3,2,1.7,1.5 M$_\odot$ (solid lines). The arrow is the
reddening vector for E(J-K)=0.4}
\end{figure}

\subsection {The extinction}
  
Using the catalog of stars in the BV pass-bands
in NGC 1893 given  Massey et al. (1995) we  identify  about 57 stars
brighter than V$\sim 18$ (or K $\sim 14$) in common with our 
study.  From the  (V-K)-(J-K) plot shown in Fig \ref{n1893vjk} we derive
the  de-reddened CMD for all the 38 stars located at the right of the
main sequence loci (see discussion in section 3.3). The resulting CMD is presented in
Fig \ref{n1893isored}. We find an  E(J-K) ranging from 0.15 to 0.5
inside the whole region. This is in agreement with
 Massey et al who find E(B-V) $ \sim $ 0.4-0.7
(corresponding to E(J-K) $\sim 0.2-0.36$)  for the
candidate members. Assuming their determination,
we select, as candidate members, objects having
 E(J-K)= 0.2-0.4. The de-reddened main sequence location is
in reasonable agreement with a 4 Myr isochrone.
We point out however, that selecting members on the basis of the reddening
is probably not very effective in regions where gas and dust
are present. The extinction is expected to be extremely  variable.

The whole observed
region is embedded in a  large cloud associated to the cluster
 (Leisawitz et al 1989)
with  CO emission. A comparison with a Digitized Sky Survey (DSS)
image allows us
to identify dark dusty clumps (see Fig.\ref{image1}).
We roughly 
define two main dark regions: $\alpha < 5^h 22^m 30^s$ and $ \delta <  33^0 22^\prime $ and $\alpha < 5^h 22^m 40^s$ and $ \delta >  33^0 24^\prime $.
 Fig.\ref{cmd1893clump} presents the CMD of these dark
clump regions where a 4$\times 10^6$ yr isochrone is
plotted.
 While in the visual pass-bands these clumps
are sparsely populated, in the JK CMD they appear to be
populated by  objects fainter than K $\sim 11$.  
Isochrones fitting to the blue edge of the main sequence allows
us to find the minimum reddening inside the region.
We derive E(J-K)
 as high as 0.45-0.55 inside the dark clumps, slightly higher than the mean reddening derived outside the dark clumps.

We compare our determination of the reddening with the 
maps by Mendez \& Van Altena (1998).
They find at the distance of 4.3 Kpc, corresponding to the de-reddened
 distance modulus of 13.20, a value of E(B-V)=0.43$\pm 0.23$ 
(or E(J-K)=0.26 $\pm 0.16$), slightly lower than, but
in reasonable agreement with
 the reddening given by NGC 1893.

\subsection{ Pre-main sequence candidates}

\begin{figure}
\centerline{\psfig{file=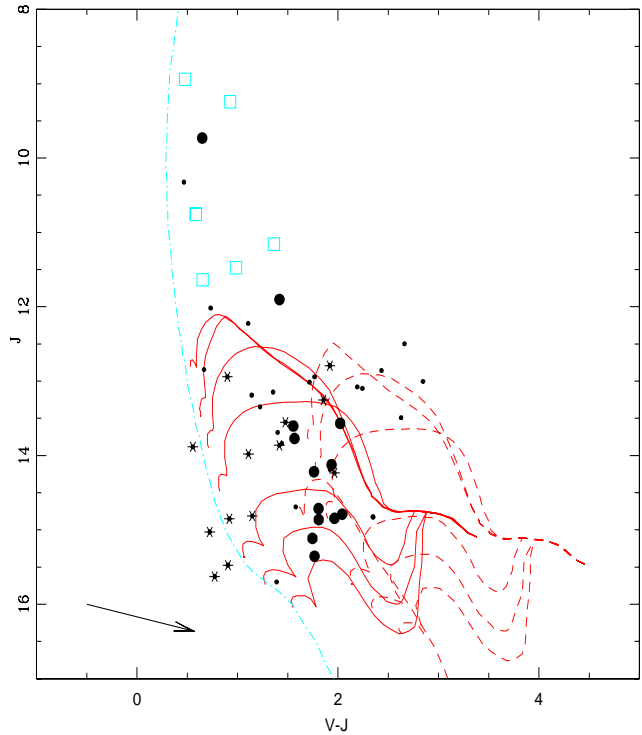 ,width=9cm,height=10cm}}
\caption{\label{n1893pre} J-(V-J) plot for 57 stars in NGC 1893. 
The dashed-dotted lines show  4 Myr isochrones having
E(J-K)=0.2.
The lines show the loci of
PMS stars by Bernasconi (1996) having  E(J-K)=0.2 (solid line) and 0.4 (dashed line)
and  masses 1.5,1.7,2,3,4,5,7 M$_\odot$. Open squares indicates main sequence
stars of known membership from spectroscopy. Solid circles are suspected members derived from reddening analyis as described in the text.
 Asterisks show objects having infrared excess. The arrow indicates the
reddening vector corresponding to  E(J-K)=0.2. }
\end{figure}

Using BV photometry by  Massey et al. (1995) 
we derive 
the J-(V-J) CMD  presented in
Fig \ref{n1893pre}.

It is not possible to distinguish between
pre-main sequence (PMS) stars and field contamination on the basis of the CMD.
However, PMS stars often show  infrared excesses
due to the presence of circumstellar disks/envelopes. Emissions
from dust and gas heated by either accretion or
photospheric radiation or both,
can be evidentiated by combining optical
and near-IR photometry (see
Hillenbrand et al. 1998).
The idea is to select PMS candidates that
populate the (V-K)-(J-K)  and the (B-V)- (J-K) plots
in a region forbidden to normally reddened stars.

Theoretical simulations (Calvet et al. 1991,
Meyer et al. 1997, Hillenbrand et al. 1998) show that the effect
of circumstellar disks/envelope on the magnitude of the resulting object
is maximum in the K band, slightly lower in the J band.
B and V colours instead can be considered as purely
photospheric, unless an accretion disk is present.
In this case, also these latter filters can be affected by a 
blue excess due to the hot accretion zone (Hartigan et al. 1991).
We note that we were not able to apply corrections for reddening
for  PMS candidates 
 while using this method, since  there are
no reliable determination of the  reddening for these stars.
Moreover, the reddening changes quite unpredictably inside the 
field of view.

\begin{figure}
\centerline{\psfig{file=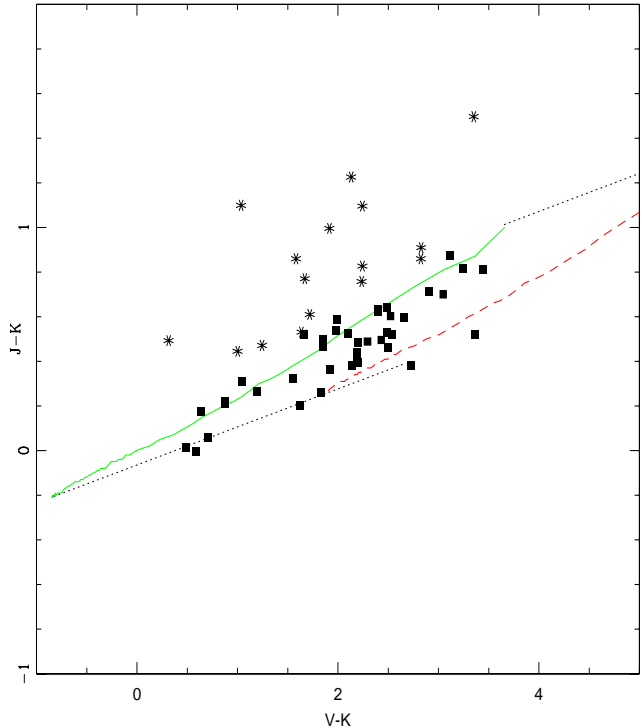,width=9cm,height=10cm}}
\caption{\label{n1893vjk} (J-K)-(V-K) plot for 57 stars in NGC 1893. The lines show the loci of
main sequence stars for E(J-K)=0 (solid line) and 0.65 (dashed line).
The reddening vectors are indicated by dotted lines. Asterisks indicate objects
suspected to be PMS candidates, as discussed in the text
}
\end{figure}

\begin{figure}
\centerline{\psfig{file=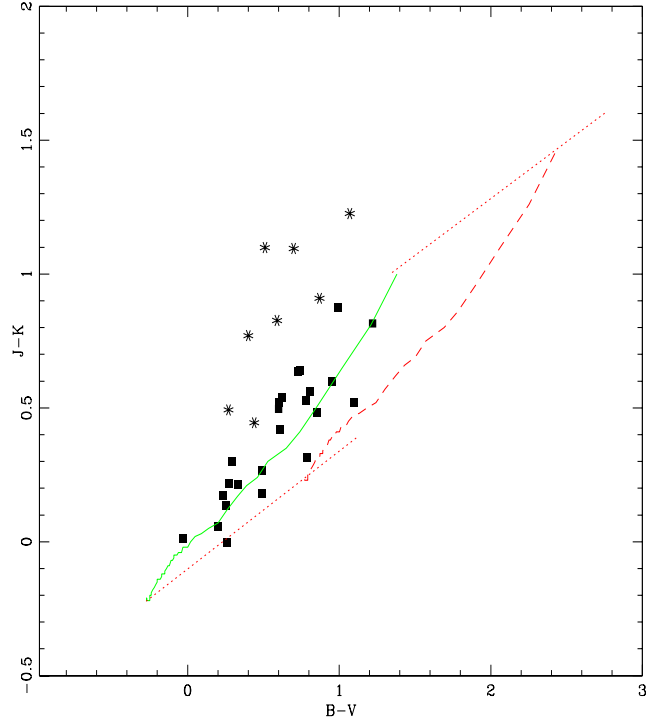,width=9cm,height=10cm}}
\caption{\label{prebvjk} (B-V)-(J-K) plot for 33 stars in NGC 1893. The lines show the loci of
main sequence stars for E(J-K)=0 (solid line) and 0.65 (dashed line).
The reddening vectors are indicated by dotted lines. Asterisks indicate objects
suspected to be PMS candidates on the basis of Fig.\ref{n1893vjk}
(see discussion in the text)}
\end{figure}

Fig.\ref{n1893vjk} presents the (J-K)-(V-K) 
diagram for 57 stars in NGC 1893.
For 33 of them  B magnitudes are available, the corresponding
 (B-V)-(J-K) diagram is 
shown Fig.~\ref{prebvjk}.
16 stars are found outside the reddening vectors
in  the (J-K)-(V-K) plot. 
To be very conservative, only stars
located 0.1 outside the reddening vectors 
are considered as candidates.
Only 8 of them
have B known magnitudes and they are located
in the (B-V)-(J-K) plot in the region forbidden
to normally reddened stars. They can be regarded
as pre-main sequence candidates.
The location of pre-main sequence stars in
the colour-colour diagrams adopted for this method obviously depend
on several parameters, including the size and geometry of the
disk/envelope, the physical properties of the dust grains,
the characteristics of the central star and more. Therefore, the
method cannot be used to assess in detail the characteristics
of the PMS candidates, but only to draw some global conclusions
on  statistical basis. 
Hillebrand et al. (1998) find that the efficiency of the 
colour-colour plots to detect PMS
T Tauri and Herbig Ae/Be stars depends on the adopted colour plot,
but is not higher than 70\%.
Additionally, we note that this method only applies to
PMS candidates with a surrounding dust disk or envelope,
and will miss all the so-called naked (or weak-line) T Tauri stars.
It is not completely clear whether these latter
represent an evolutionary phase or an independent phenomenon,
but we know that depending on the
star-forming region their incidence can be quite important
even at ages of a few million years (for instance, in the Taurus-Auriga
complex). Therefore,  we conclude that a fortiori our estimate
of the number of PMS candidates in  NGC~1893 is a lower limit.

From the Log T-Log L plane PMS tracks by Bernasconi (1996)
we calculate the BVJK magnitudes, using the tables of corrections
by Bertelli et al. (1994).
The comparison with the observations in the J-(V-J) plane
(cf Fig. \ref{n1893pre})
suggests
that the majority of the   stars having IR-excess can be 
 PMS objects located 
close to  a reddened 4 Myr isochrone (E(J-K) $\sim$ 0.2-0.4) and
having 2-4 and possibly 7 M$_\odot$ masses. 
If all the stars inside NGC~1893 are born at the same time, we expect that
objects more massive than  2-3 M$_\odot$ have already reached
 the main sequence.
Indeed,  a star having 7 M$_\odot$ mass
needs about 6.3 $\times 10^5$  to reach the main sequence, and
a 4 M$_\odot$ mass star needs 1.5  $\times 10^6$ yr.
If these  candidates are true PMS stars a large
spread in age
might be present. However, since in these passbands the reddening vector
is almost parallel to the x-axis, differential reddening can 
be  responsible for the  location of these
intermediate mass PMS candidates  in the CMD.
It cannot be excluded that some of the PMS
candidates belong to the field population. Additionally,
when combining optical and near-infrared photometry which have not
been obtained simultaneously, the expected strong variability of the sources
might  affect the results.
 Definitive identification of accreting
systems can be obtained only by mean of spectroscopic determination.

As we discussed in the previous section,
an optically obscured  clump with CO emission is located inside the observed field. Such a region might be the birthplace of young
objects. 
A comparison of the CMD of this region
with PMS tracks by Bernasconi (see Fig \ref{cmd1893clump}) suggests that at least some
of the detected stars can be PMS candidates. Since almost no
stars inside this region have BV photometry, the method of the colour-colour
plot to detect PMS candidates cannot be applied.

\begin{figure}
\centerline{\psfig{file=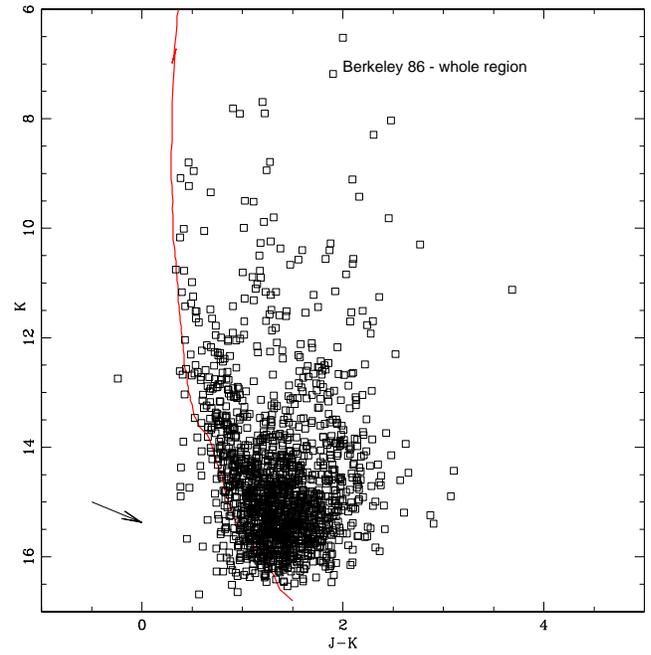,width=9cm,height=9cm}}
\caption{\label{be86iso} CMD of
Berkeley~86 (whole region) with isochrone from Padova models having
an age of 6 Myr and E(J-K)=0.5. The arrow indicates the reddening vector
corresponding  to E(J-K)=0.5}
\end{figure}

\begin{figure}
\centerline{\psfig{file=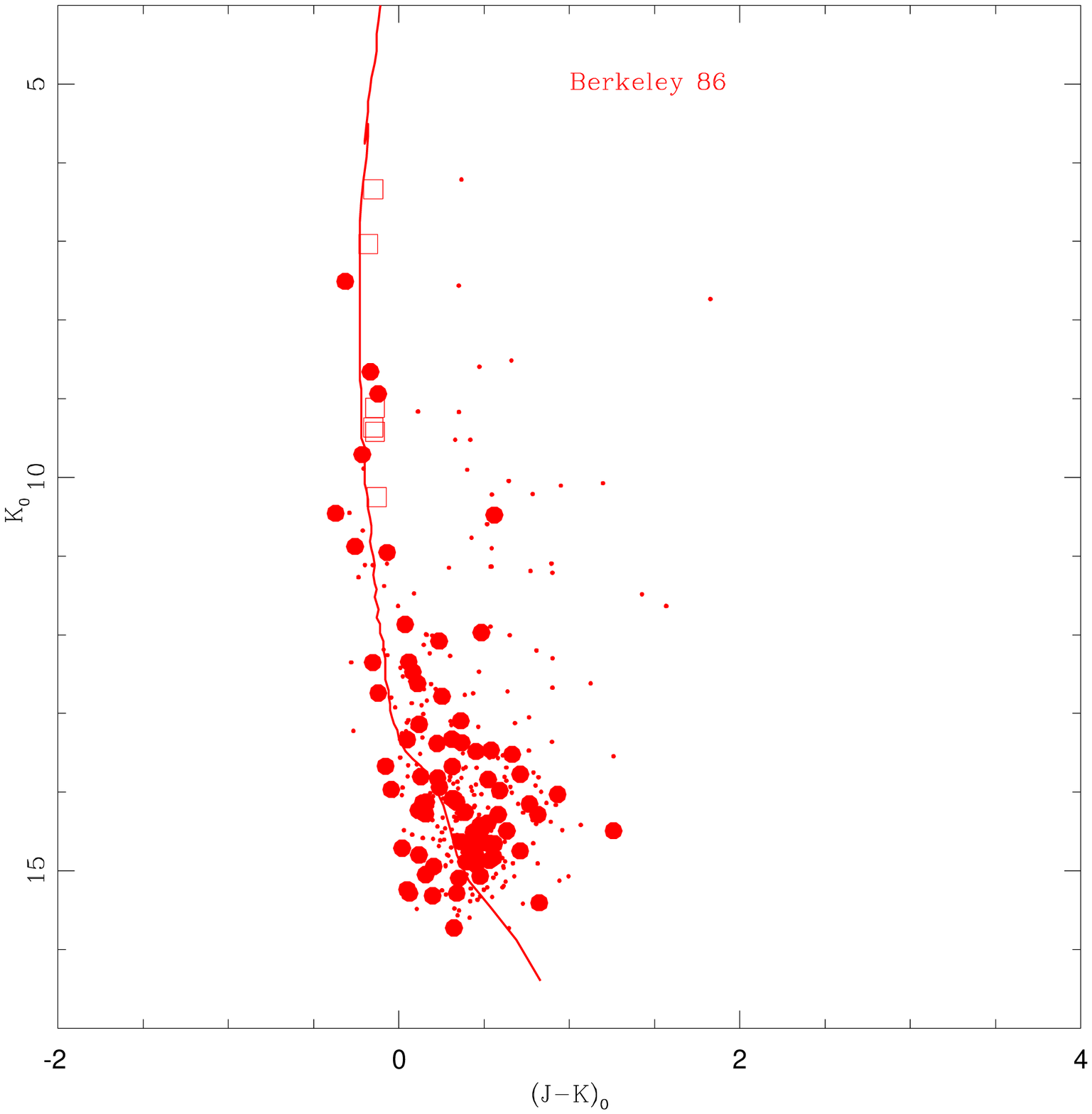,width=9cm,height=9cm}}
\caption{\label{cmdbe86red} Reddening corrected CMD of
Berkeley~86  with an isochrone from Padova models having
an age of 6 Myr. Solid circles indicate suspected members, selected
on the basis of the reddening, open squares represent members known from
spectroscopy (Massey et al 1995).}
\end{figure}

\section{Berkeley~86}

\subsection {The CMD}

The CMD  of the observed region inside  Berkeley~86
is shown in Fig.\ref{be86iso}. 
The contamination from interlopers is very high. However it is
possible to distinguish a vertical MS.  
With a distance modulus of $(m-M)_{o}$=11.1 (Massey et al 1995)
an  isochrone of 6 Myr and  E(J-K)=0.5 seems to reproduce
the main sequence location.

We identify about 340 stars in common with Deeg \& Ninkov (1996)
photometry for which
UBVRI magnitudes are available. Using their estimate of the
E(B-V) for each stars we derive  the corresponding E(J-K), following
the relation E(J-K)/E(B-V)=0.52 (Cardelli et al 1989).
The final K-(J-K) CMD reddening corrected is presented in
Fig \ref{cmdbe86red}. E(J-K) is found ranging from 0.45 to 0.64.
The relative frequency of objects inside the field with the  E(J-K)
 is peaked at  E(J-K) $ \sim$ 0.5. 
The suspected members are selected on the basis
of the reddening, E(J-K) ranging from 0.48 to 0.52, roughly
corresponding to 1 $\sigma$ in the color excess distribution.
As in the case of NGC~1893 we point out that a selection of the members
on the basis of the reddening gives only a rough indication
 in regions where
dust and gas are present, since the extinction might be highly variable.

The reddening maps by  Mendez \& van Altena  (1998) give E(B-V)=0.96 or E(J-K) $\sim $ 0.41
at the distance of 1700 pc, where Berkeley~86 is located. The maximum expected reddening inside the region is E(B-V)=2.3 or  E(J-K)=0.99. This is in good
agreement with our determinations, considering that the mean error
on the  Mendez \& van Altena maps is about  $\delta$ E(J-K)=0.16.

\subsection{Pre-main sequence candidates}

\begin{figure}
\centerline{\psfig{file=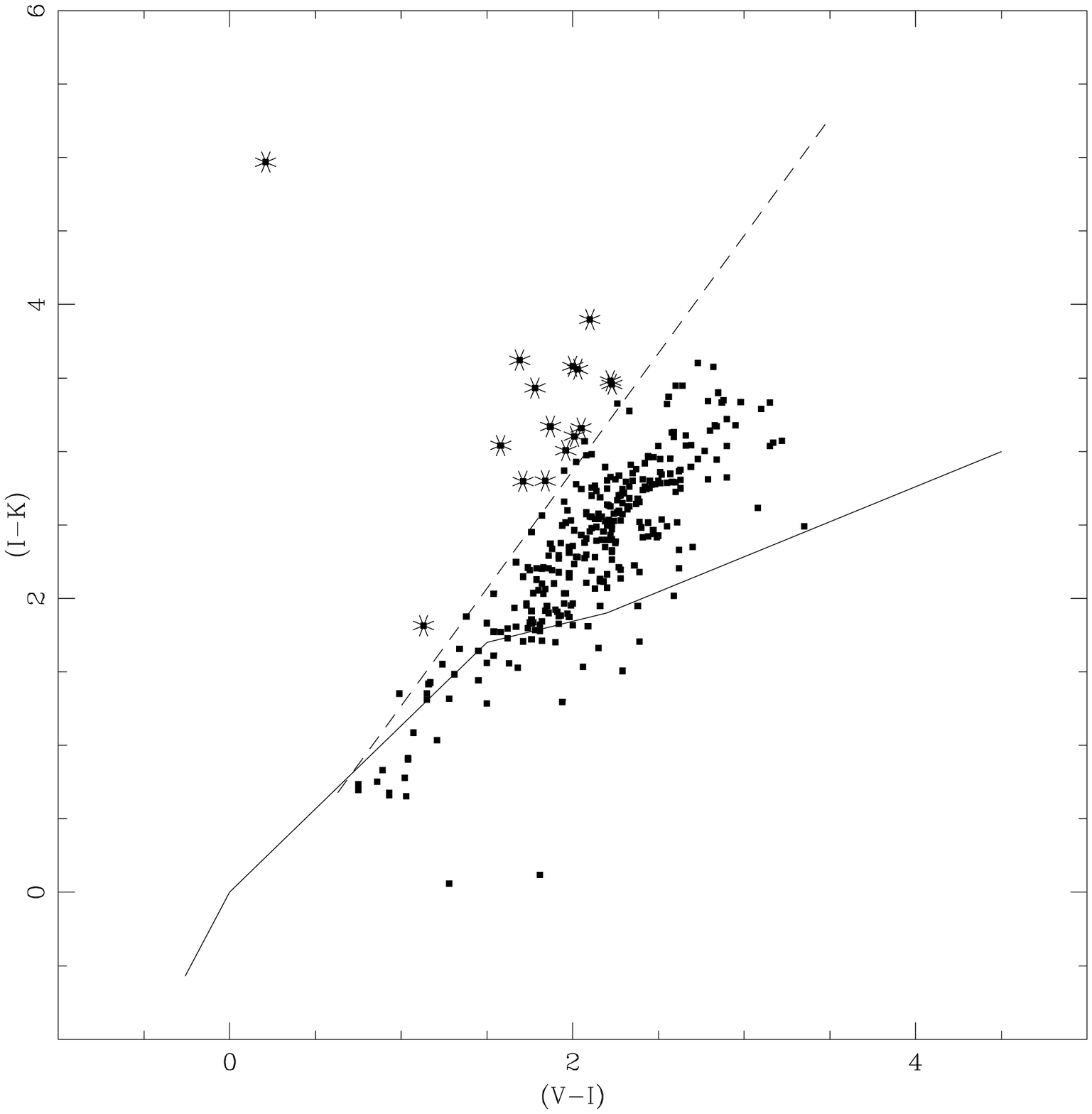,width=9cm,height=10cm}}
\caption{\label{be86vjk}
(V-I)-(I-K) plot for about 340 stars in Berkeley~86. The solid line give the loci
of the main sequence and giant stars  having reddening E(J-K)=0.
 Reddening vector is indicated by a dashed line. Asterisks show the
candidate pre-main sequence stars.}
\end{figure}

\begin{figure}
\centerline{\psfig{file=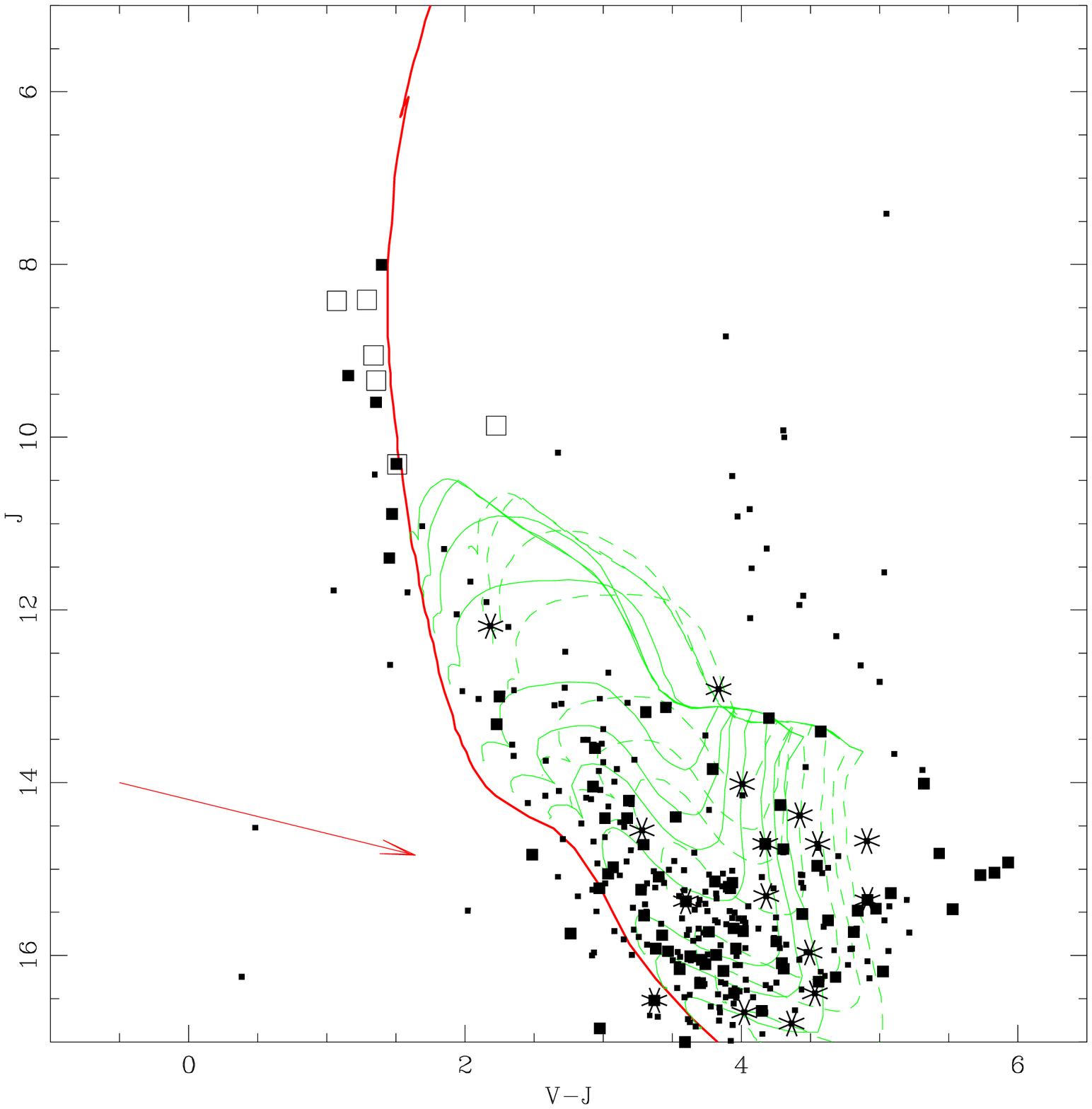,width=9cm,height=10cm}}
\caption{\label{be86vk} J-(V-J) plot for about 340 stars in Berkeley~86. The
stark solid line represents a 6 Myr isochrone
 having E(J-K)=0.5. Stars indicate
pre-main sequence candidates. Open squares show members known from
spectroscopy. Solid squares present suspected members selected
on the basis of the reddening (see text).
 Dashed lines  are pre-main sequence
tracks by Bernasconi, for masses 0.8,0.9,1.0,1.2,1.5,1.7,2,3,4,5,7
and reddening   E(J-K)= 0.6 and solid lines are the analogous
for E(J-K)= 0.5.
 The arrow is the reddening vector  corresponding
to E(J-K)= 0.5.}
\end{figure}

Coupling Deeg \& Ninkov photometry with  our JK magnitudes,
we discuss the presence of candidate pre-main sequence stars.
Since I magnitudes are available, we make use of the
(V-I)-(I-K) plot. Hillenbrand et al have shown that this color
combination is specially effective in identifying stars with infrared
excess.
Fig.\ref{be86vjk} presents the (V-I)-(I-K) plot for the stars in common with 
Deeg \& Ninkov (1996). 16 stars seem to be good 
pre-main sequence candidates. In Fig.\ref{be86vk}  their location in
the  J-(V-J) CMD is presented. A comparison with the pre-main sequence
tracks shows that the majority of these candidates might have low masses, between
 1.7 and 0.8 M$_\odot$. Only two candidates are found in the region occupied
by higher mass stars. We cannot exclude however that these stars belong
to the background population. 
Since the time a 1.7 M$_\odot$ star need to reach the main sequence 
is longer than the age of Berkeley~86 (about 1.5 $\times 10^7$ yr, Bernasconi 1996) no need
of a large age spread is found inside the population of Berkeley~86.

\section{Conclusions}
We have presented J and K near-infrared photometry down to
a limiting magnitude of K $\sim$17 and 16.5 for the two very young,
highly obscured open clusters NGC~1893 and Berkeley~86, respectively.
For Berkeley~86 near-infrared photometry was not available before our
study, while in the case of NGC~1893 previous near-infrared photometric
survey 
 only reached K=13. 
We confirm previous determinations of the age,
distance and reddening of these clusters.
Coupling J-K data with Johnson  photometry we discuss
the presence of stars with IR excess, as pre-main sequence candidates.
Several good
candidates are identified in both clusters, confirming the young
age of these objects. Spectroscopic measurements are
requested to definitively identify pre-main sequence objects.

\begin{acknowledgements}
The authors are thankful to the referee C. Dougados for many
useful comments.
This research has been sponsored by the Italian Ministry
of University and Research, and by the Italian Space Agency.

\end{acknowledgements}

{}

\end{document}